# Large Seebeck Effect by Charge-Mobility Engineering

[1]Peijie Sun, [1]Beipei Wei, [1]Jiahao Zhang, [2]Jan M. Tomczak, [3,4]A. M. Strydom,
[5]M. Søndergaard, [5]Bo B. Iversen, and [4,1]Frank Steglich

[1] *Beijing National Laboratory for Condensed Matter Physics, Institute of Physics, Chinese Academy of Sciences, Beijing 100190, China*

[2] *Institute of Solid State Physics, Vienna University of Technology, A-1040 Vienna, Austria*

[3] *Highly Correlated Matter Research Group, Physics Department, University of Johannesburg, Auckland Park 2006, South Africa*

[4] *Max Planck Institute for Chemical Physics of Solids, 01187 Dresden, Germany*

[5] *Department of Chemistry, University of Aarhus, DK-8000 Aarhus C, Denmark*

(Dated: February 20, 2015)

Email: pjsun@iphy.ac.cn

PACS numbers: *Valid PACS appear here*

**The Seebeck effect describes the generation of an electric potential in a conducting solid exposed to a temperature gradient. Besides fundamental relevance in solid state physics, it serves as a key quantity to determine the performance of functional thermoelectric materials. In most cases, it is dominated by an energy-dependent electronic density of states at the Fermi level, in line with the prevalent efforts toward superior thermoelectrics through the engineering of electronic structure [1–3]. Here, we demonstrate an alternative source for the Seebeck effect based on charge-carrier relaxation: A charge mobility that changes rapidly with temperature can result in a sizeable addition to the Seebeck coefficient. This new Seebeck source is demonstrated explicitly for Ni-doped $CoSb_3$, where a dramatic mobility change occurs due to the crossover between two different charge-relaxation regimes. Our findings unveil the origin of pronounced features in the Seebeck coefficient of many other elusive materials characterized by a significant mobility mismatch. As the physical origin for the latter can vary greatly, our proposal provides a unifying framework for the understanding of a large panoply of thermoelectric phenomena. When utilized appropriately, this effect can also provide a novel route to the design of improved thermoelectric materials for applications in solid-state cooling or power generation.**

Even though no fundamental constraints other than the Carnot efficiency are known to limit the potential efficiency of thermoelectric materials [2], the dimensionless figure of merit $ZT = T \cdot S^2\sigma/\kappa$ ($T$ is the absolute temperature, $S$ the Seebeck coefficient, $\sigma$ the electrical conductivity and $\kappa$ the thermal conductivity) in current materials hovers at values around 1, which are still inadequate for widespread applications [3,4]. Efforts to optimize $ZT$ in general fall into two categories: "phonon engineering" and "electronic-structure engineering". The most successful concept in this context is that of a phonon glass-electron crystal [5]. Here one attempts to reduce the thermal conductivity by incorporating intrinsic or extrinsic phonon scatterers while keeping the electronic properties unchanged [6]. Conversely, "electronic-structure engineering" [1–3] pursues a favorable electronic dispersion at the Fermi level, $\epsilon_F$, in order to achieve a higher thermoelectric power factor ($S^2\sigma$). Optimization of $ZT$ through these concepts appears to be limited by the interdependence of the three relevant transport coefficients ($S$, $\sigma$, $\kappa$). In some specific systems more exotic physical phenomena such as spin degeneracy and electronic correlations are also highly relevant to the Seebeck effect [7,8], however, their impact is less transparent in practical material design.

The Seebeck coefficient of a conducting solid is generally induced by the asymmetry of the electronic density of states (DOS) $N(\epsilon)$ at the Fermi level $\epsilon_F$, as illustrated in Fig. 1a. There, a temperature gradient $\Delta T_x$ along the sample leads to a slight gradient in $\epsilon_F$ [9]. Due to the energy dispersive $N(\epsilon)$ and the Fermi function $f(\epsilon)$, which both determine the charge-carrier density at $\epsilon_F$, $n(\epsilon_F)=N(\epsilon_F)f(\epsilon_F)$, a net diffusion of electrons occurs along the sample. The electron diffusion is eventually impeded by a retarding electric potential ($V_x$), leading to $S_N = V_x/|\Delta T_x|$ at equilibrium. We denote this conventional contribution to the Seebeck effect as $S_N$. This term also explains Seebeck's original observation of an electric potential across the junction of two metallic wires (cf. Fig. 1b). Here, the difference of the Fermi energies of the two metals simply sets up an artificial energy dependence of $N(\epsilon)$.

Below, we demonstrate a new source for the Seebeck effect, $S_\tau$, based on a charge mobility, $\mu_H(T)$, that changes rapidly with temperature. As illustrated in Fig. 1c, a large derivative $d\mu_H(T)/dT$ can generate a sizeable component $S_\tau$ adding to the conventional Seebeck coefficient $S_N$. As we shall see below, this scenario unifies the thermoelectric manifestations of various physical phenomena that lead to a significant gradient of charge relaxation processes. The contribution $S_\tau$ can easily be distinguished from $S_N$ by measuring the transverse thermoelectric signal (i.e., the Nernst coefficient $v$) in the presence of a magnetic field perpendicular to both $S$ and $v$. In materials with rapidly changing mobility, the Lorenz force affecting the slow and fast carriers of the cold and hot ends (or vice versa) will not fully compensate, leading to a sizeable value of $v$ that is dependent on $S_\tau$ (Fig. 1c). In cases with an insignificant thermal variation of $\mu_H(T)$ (Fig. 1a), the so-called Sondheimer cancellation removes the transverse potential since the Lorenz forces on the forward and backward flows of the electron current exactly compensate [10,11]. In analogy to the case shown in Fig. 1b, a proper engineering of the charge mobility e.g. by artificially fabricating a junction between two conducting solids of very different mobilities, promises to further enhance the Seebeck effect, cf. Fig. 1d.

The idea of using mobility gradients to enhance the Seebeck effect emerged from re-assessing the recent thermoelectric investigations on a specific material class of heavy fermions by two of the authors [12]. There, an enhanced Seebeck coefficient was attributed to the asymmetric Kondo scattering processes of conduction electrons from localized 4$f$ electrons. Characterizing the energy-dependent charge relaxation time $\tau(\epsilon)$ due to the Kondo effect, a significant Nernst coefficient $v(T)$ was observed and found to account for the enhanced Seebeck coefficient after renormalization by $\mu_H$. This implies an inactivation of the Sondheimer cancellation, accounting for a sizeable $S_\tau \approx S$ in heavy fermions. Conceptually different from the local Kondo process, charge relaxation time in a conventional solid is usually dominated by weakly energy-dependent scattering events, which, however may exhibit a large temperature dependence due to, e.g., a crossover between different relaxation regimes. As far as the thermal transport is concerned, an enhanced gradient of $\tau$ with respect to temperature is physically identical to an energy-dependent $\tau(\epsilon)$ as evidenced by, e.g., the Kondo effect. Notably, the former situation is much more transparent: Regardless of the origin - we will discuss several possibilities below -, sufficiently different charge mobilities at the hot and cold ends of a material will generate an electrical potential analogous to modifying the DOS of an energy-dependent $N(\epsilon)$ (cf. Figs. 1a and 1c).

In the following, we will demonstrate the pivotal influence of the above mechanism for a weakly Ni-doped skutterudite $CoSb_3$. As shown in Fig. 2 (see also ref. [13]), $S(T)$ of $Co_{0.999}Ni_{0.001}Sb_3$ is negative and only weakly temperature dependent above 50 K. Further cooling of the temperature leads to a dramatic sign change of $S(T)$ and a pronounced positive peak of 110 µV/K at $T \approx 20$ K. Here we highlight the opposite signs of $S(T)$ and $R_H(T)$ below 30 K, the latter being negative in the whole temperature range investigated (cf. Fig. 3a). Phonon-drag effects can safely be excluded as the origin of the extraordinary $S(T)$ peak at 20 K due to its absence in undoped

CoSb$_3$ [13]. In view of the opposite signs of $S(T)$ and $R_H(T)$ and in order to account for the unusual peak in the former, an intuitive approach involves a two-band model, with charge carriers of different signs being involved. This is, however, in contradiction to the Hall resistivity data that hint to two *electron-like* bands (cf. Supplementary Information (SI)), whose influences furthermore only coexist in a very limited temperature region (25 −40 K, hatched in Fig. 2). Near 20 K, the one-band nature is restored. These observations imply that, while two-band effects are involved in a limited temperature interval (cf. Figs. 2 and 3a), they are not responsible for the *positive S(T)* peak. Another likely multiband regime below $T \approx 7$ K (cf. Fig. 3a), is beyond the scope of this investigation. In view of the sublinear $\rho_H(B)$ above 10 K (SI), we proceed our discussion in the framework of a one-band picture. This applies at least to the low magnetic field region ($B \leq 2$ T) where we have measured $v(T)$ and estimated $R_H(T)$.

The occurrence of a peak in $R_H(T)$ and a shoulder in $\rho(T)$ at $T \approx 40$ K (cf. Fig. 3a and 3b) supports the notion of an *electron-like* shallow impurity level that becomes active below 40 K, due to the surplus electrons of Ni atoms (SI). Applying the thermal activation law $\rho = \rho_0 \exp(E_a/k_B T)$ to the $T$ range 40−110 K above the shoulder (region I, Fig.3b), the activation energy $E_a$ is estimated to be 207 K. This is associated with either an intrinsic electronic band or a deeper donor level [13]: As revealed in Fig. 3a, these states are characterized by a high charge mobility. Due to the narrow temperature window, we have not estimated the activation energy for region II (15−40 K). Instead, we stress that the data below ~15 K (region III) can be well described by the variable range hopping (VRH) model, expected for conducting states of weak localization. The observation of negative values of magnetoresistance and a characteristic dependence of $\log \rho \sim (1/T)^{1/4}$ in regime III lend strong support to this proposition (Fig. 3 and inset; ref.[13]). Consequently, a pronounced change of $\mu_H(T)$ connecting a region with high mobility (intrinsic or deeper-donor derived) charge carriers to a low-mobility VRH conduction occurs due to the onset of a shallow impurity level. The pronounced change of the mobility, following $\mu_H(T) \sim T^7$ between 40 and 10 K (Fig. 3a), dwarfs the characteristics of common charge scattering processes, e.g., the $T^{3/2}$ dependence expected for ionized impurity scattering.

What are the consequences for the Nernst response? In a nonmagnetic, nonsuperconducting one-band system [10, 14],

$$v = -\frac{\pi^2}{3} \frac{k_B^2 T}{B|e|} \frac{\partial \tan \theta_H}{\partial \varepsilon}\bigg|_{\varepsilon = \varepsilon_F} \quad (1)$$

Here the energy derivative of the tangent of the Hall angle can be expressed by using either $\mu_H$ or $\tau$

because $\tan\theta_H = eB\tau/m^* = \mu_H B$, where $m^*$ denotes the effective mass of the relevant charge carriers. Apparently, $v$ is sensitive to any charge relaxation process that is asymmetric with respect to $\epsilon$. However, the asymmetry associated with ordinary scattering events is typically rather weak, and representable by a power-law dependence of $\tau(\epsilon) \sim \epsilon^r$ with $|r| \approx 1$ for, e.g., electron scattering by acoustic phonons. Replacing $1/\partial\epsilon$ in Eq. 1 by $1/k_B\partial T$ [9], one immediately recognizes that an enhanced gradient of $\mu_H$ with respect to $T$ similarly can supply a finite Nernst coefficient $v = AT d\mu_H/dT$, where $A = -(\pi^2/3) k_B/|e|$. As can be seen in Fig. 4, the pronounced $v(T)$ peak observed slightly below 40 K for $Co_{0.999}Ni_{0.001}Sb_3$ follows this prediction *quantitatively*, underlining the existence of distinctly different charge mobilities at the two ends of the sample when exposed to a temperature gradient (cf. Figs. 1c and 3a). Given that above 40 K $\mu_H(T)$ is only weakly dependent on temperature, one expects $v(T) \propto \mu_H(T)$ ( $d\mu_H/dT$ is approximated to first order as $\mu_H/T$, cf. Ref. [10]). This is also confirmed in Fig. 4 and at $T > 40$ K, $v(T)$ scales well with $\mu_H(T)$ (red dashed line).

We now scrutinize how the large mobility gradient impacts upon the Seebeck effect. The Mott expression relates the Seebeck coefficient to the logarithmic energy derivatives of $N(\epsilon)$ and $\tau(\epsilon)$ at the Fermi energy,

$$S = -\frac{\pi^2}{3}\frac{k_B^2 T}{e}\left[\frac{\partial \ln \tau}{\partial \varepsilon} + \frac{\partial \ln N}{\partial \varepsilon}\right]_{\varepsilon_F} = S_\tau + S_N. \qquad (2)$$

Even though $S_\tau$ due to asymmetric charge relaxation has been ignored in most thermoelectric explorations, it can in fact be the dominating term in specific cases such as heavy fermion systems, as was realized recently [12]. As discussed in the SI and refs. [11, 12], the Nernst coefficient is physically linked to $S_\tau$ by $v \cdot B = \pm S_\tau \tan\theta_H$ (or, $v = \pm S_\tau \mu_H$), since $\tan\theta_H \propto \tau$ in a conventional one-band solid. An additional $S_\tau = -v/\mu_H$ is therefore expected to occur in $Co_{0.999}Ni_{0.001}Sb_3$ due to the abrupt mobility change. Here the negative sign refers to the electron-like charge carriers. Remarkably, the ratio $-v/\mu_H$ can indeed account for the abrupt crossover and the *positive S(T)* peak emerging below 50 K (dashed line in Fig. 2). This offers compelling evidence that the rapidly changing mobility is underlying this extraordinary thermoelectric response.

To substantiate this assessment also from the theoretical point of view, we have computed the Seebeck coefficient based on state-of-the-art *one-particle* electronic structure calculations, assuming a *constant* scattering rate $\tau$. While by construction missing out on the mobility-driven effects, this procedure is expected to qualitatively capture all intrinsic density-of-states effects, and therewith in particular the contribution $S_N$ to the Seebeck coefficient. This expectation is indeed fulfilled: The theoretical $S_N$ of $Co_{0.999}Ni_{0.001}Sb_3$, which is negative for all temperatures, approaches the experimental result at only high temperatures where $S_\tau$ is expected to be less relevant (cf. Fig. S3c in SI).

In view of both the very similar $\mu_H(T)$ profiles and the competing signs of $S(T)$ and $R_H(T)$ at low temperatures, the scenario described above will be able to explain the positive peak in $S(T)$ observed also for other doping levels of $Co_{1-x}Ni_xCo_3$ reported in ref. [13]. One may notice that, in the nominally pure $CoSb_3$, a similar mobility gradient exists with, however, $R_H(T)$ being positive. The opposite sign of the charge carriers in pure $CoSb_3$ relative to that of the Ni-doped $CoSb_3$ will naturally invert the sign of the corresponding $S_\tau$. This may account for the abrupt drop of $S(T)$ in $CoSb_3$ from positive to small negative values below $c.100$ K (cf. Fig.5 of ref. [13]).

In Table 1, we summarize the expected signs for the mobility-driven Seebeck effect ($S_\tau$) and the Nernst coefficient ($v$) based on Eqs. 1 and 2. The signs are solely determined by the direction of the mobility gradient and the type of the charge carriers. Clearly, a negative $d\mu_H/dT$ is technically more attractive since in this case, a positive (negative) $S_\tau$ would add to a $p$-type ($n$-type) transport, leading to an overall enhancement of the Seebeck coefficient.

The significance of our findings is straightforward, yet wide-ranging: An additional Seebeck coefficient $S_\tau$ is generated whenever a temperature-dependent $\mu_H$ exists. However, since $S_\tau = \pm v/\mu_H$, the mobility-driven Seebeck effect will be sizeable or even dominating only if the gradient $d\mu_H/dT$ is huge, but not $\mu_H$ itself. This concept is far more general than what has been discussed for heavy fermions based on the Kondo effect. We can think of a variety of materials whose thermoelectric responses conform to our unifying scenario: For example, in $Cu_2Se$, a structural phase transition occurs at approximately 400 K, where $\mu_H$ changes abruptly with temperature due to critical scattering. Correspondingly, an unconventional Seebeck coefficient contributing largely to the enhanced $ZT$ values has indeed been observed and attributed to abnormal scattering processes concomitant to the phase transition [15]. Importantly, the sign of the additional Seebeck coefficient indeed follows our prediction shown in Table 1. Another example that deserves attention is $AgBiSe_2$ [16], which develops a large negative $S(T)$ maximum out of $p$-type transport at around 580 K. This feature concurs with a pronounced change of the electrical conductivity, where a dramatic change of the mobility is expected [17]. Our scenario appears to be active also in the recently discovered heavy-fermion compound $CeRu_2Al_{10}$ [18], where a sharp negative gradient of $\mu_H(T)$ is found at $T \approx 20$ K, along with an anomalous $S(T)$ peak. These observations can indeed be described within the current framework, as shown in SI. More examples may be expected/found in compounds where a phase transition or crossover between two regimes of significantly different charge-relaxation mechanisms, e.g., associated with Mott or Anderson localization, takes place. One may also envision engineering the charge mobility of a heterogeneous semiconductor along a certain direction by, e.g., controlled doping. Alternatively, one could fabricate an artificial junction from two, preferably already functionally well-graded, thermoelectric materials with very different mobilities to further improve the performance. By disentangling the impact of the gradients of the one-particle DOS and the charge mobility (or life time) onto thermoelectricity, our findings will also help answering fundamental questions in more complex systems. This pertains in particular to electron-correlated materials, in which the two gradients are convoluted in the many-body spectral function. Further investigations to explore, as well as exploit, the overarching framework of mobility gradients are eagerly called for, both from the fundamental and technological points of view.

TABLE I: The expected sign of the mobility-driven Nernst coefficient $v$ and Seebeck effect $S_\tau$ for the different combinations of the signs of both the charge carrier and $d\mu_H/dT$. The symbols + and − denote positive and negative values, respectively.

| $d\mu_H/dT$ \ charge carrier | hole | electron |
|---|---|---|
| + | $v(-); S_\tau(-)$ | $v(-); S_\tau(+)$ |
| − | $v(+); S_\tau(+)$ | $v(+); S_\tau(-)$ |

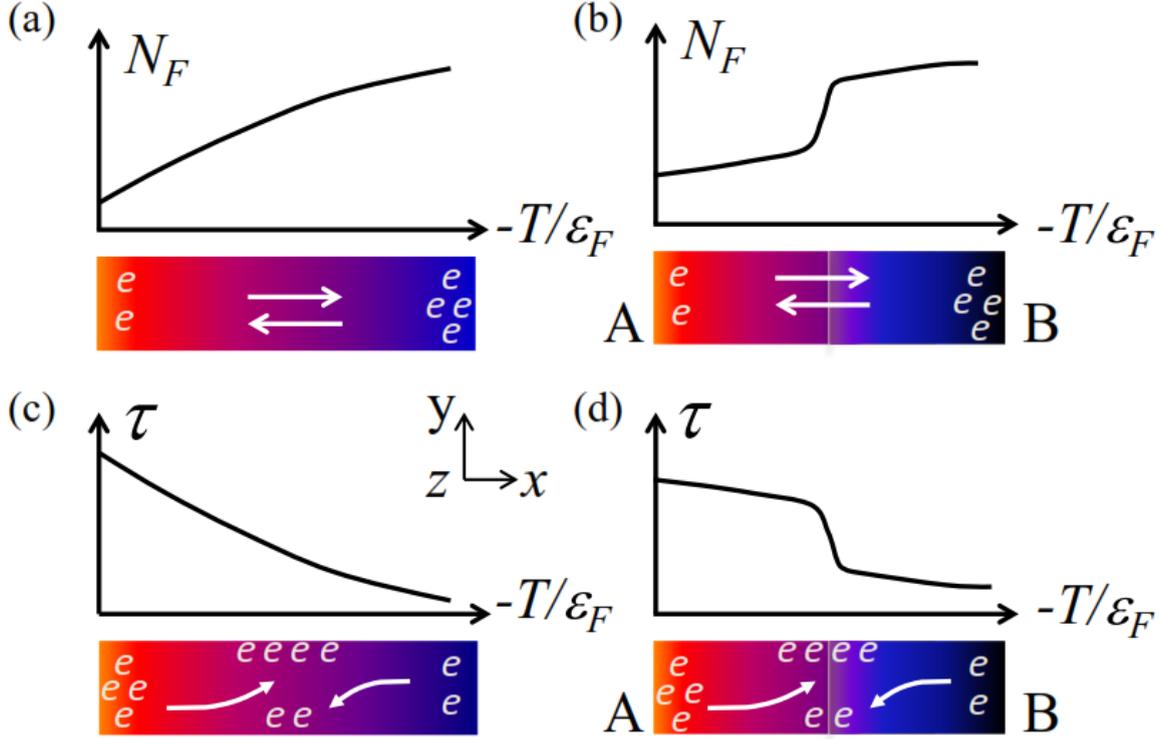

FIG.1: The Seebeck effect derived from different asymmetries of charge carriers at the Fermi level. (a) A conducting solid with a significant energy-dependent DOS. (b) A junction of conducting solids A and B with different DOS, which is the situation where the Seebeck effect was originally discovered. (c) A conducting solid with a steep energy dependence of the electron relaxation time $\tau$. (d) A junction between two conducting solids of significantly different $\tau$. The vertical axis denotes either $N(\epsilon)$ or $\tau(\epsilon)$ at the Fermi level. The horizontal axis denotes temperature, or equivalently, the Fermi energy, due to their correlation [9]. Note that scenarios (c) and (d) both produce a $\tau$ mismatch, which we exploit towards an enhanced Seebeck effect. When applying a magnetic field along the $z$ direction, only in these two cases, transverse electric potential along the $y$ direction (the Nernst effect) can be expected. In the case (a) and (b), such a signal is fully compensated due to the Sondheimer cancellation (see text).

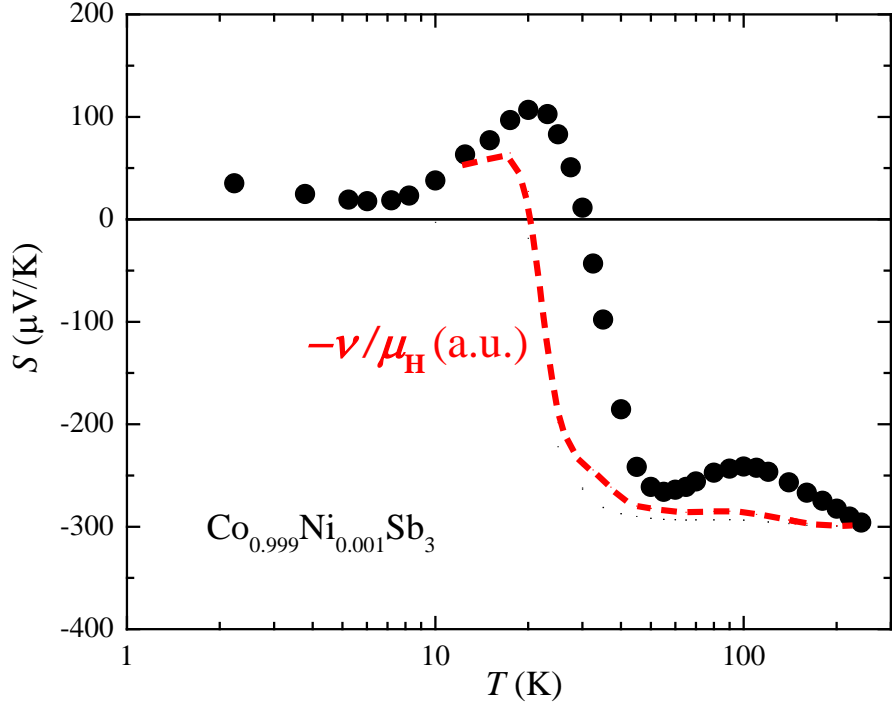

FIG. 2: The Seebeck coefficient, $S(T)$, of $Co_{0.999}Ni_{0.001}Sb_3$. The measured $S(T)$ is compared with calculated values of $\nu/\mu_H$, i.e., the expected Seebeck contribution, $S_\tau$, derived from the mobility gradient, see text. Note that the positive peak developed below 50 K due to the *n*-type charge carriers can be well reproduced by this ratio. The hatched area indicates the temperature window where two *electron-like* bands are involved, and these cannot describe the occurrence of the *positive* $S(T)$ peak.

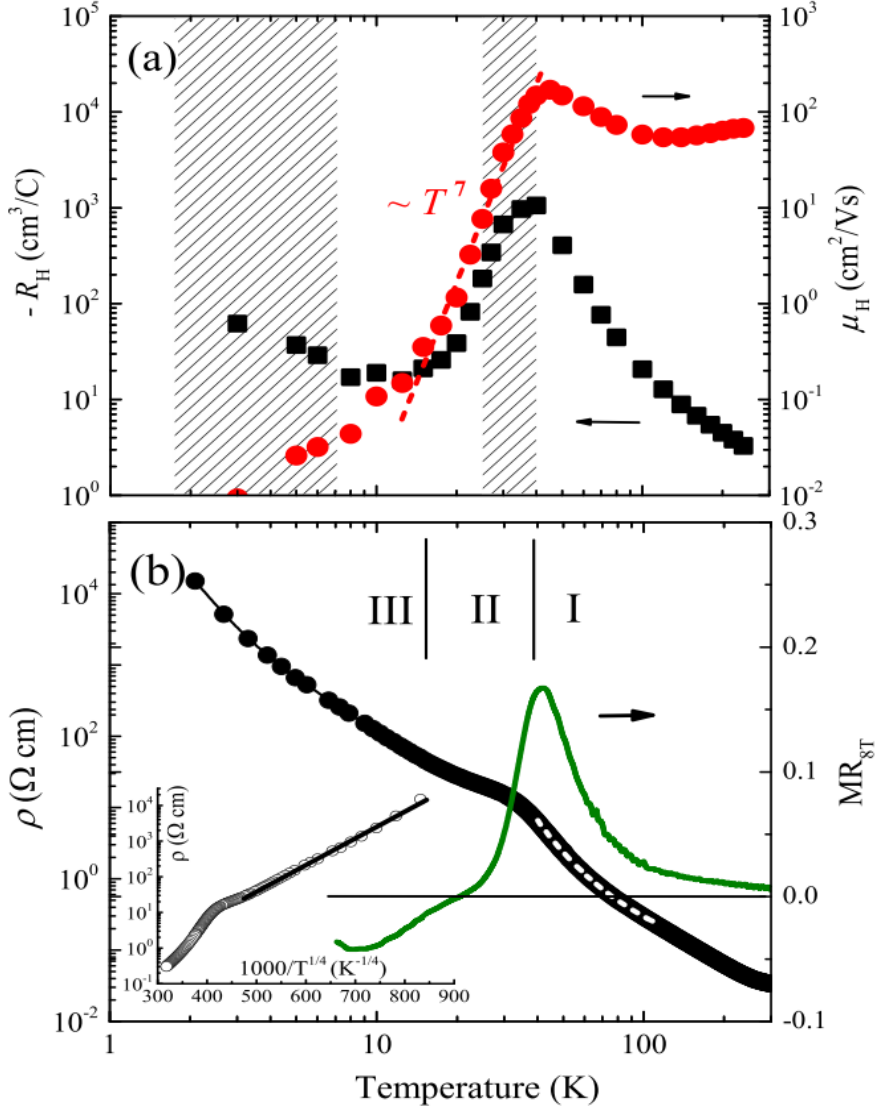

FIG. 3: Electrical transport properties of $Co_{0.999}Ni_{0.001}Sb_3$. (a) Hall coefficient $R_H(T)$ and Hall mobility $\mu_H(T)$, defined as $|R_H(T)|/\rho(T)$. The dashed line indicates an approximate $T^7$ dependence of $\mu_H(T)$. The hatchings indicate the regions where multiband effects are present. The one at around 30 K involves two electron-like bands and the one below 7 K is complex and beyond the scope of this work. (b) Electrical resistivity $\rho(T)$ and magnetoresistance $MR(T) = [\rho_B(T) - \rho_0]/\rho_0$. The latter quantity was measured in a magnetic field $B = 8$ T. Three different regions (I, II, III) are indicated, of which region I is characterized by high mobility, intrinsic or deeper-donor derived conductance and region III by low mobility variable range hopping. The dashed line on top of the $\rho(T)$ curve represents a fit by the thermal activation law. Inset: Electrical resistivity plotted as $\log \rho$ vs $T^{-1/4}$. A linear dependence, as well as the negative $MR(T)$ below about 15 K is characteristic of VRH conduction.

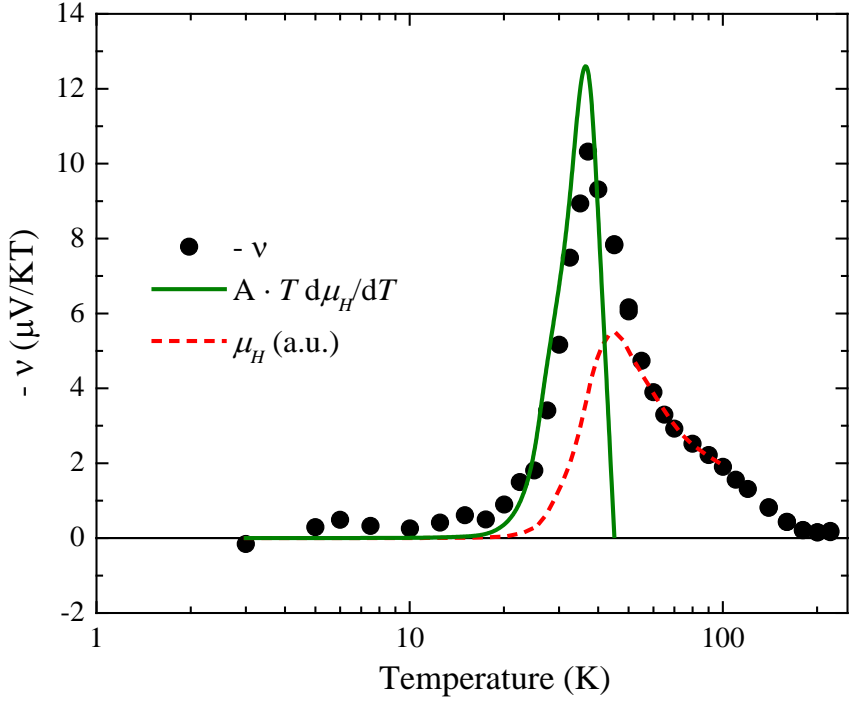

FIG. 4: The Nernst coefficient of $Co_{0.999}Ni_{0.001}Sb_3$. The green solid line represents $ATd\mu_H/dT$, the expected contribution to $\nu(T)$ from the dramatic charge mobility change (cf. Eq. 1), where the prefactor $A=-(\pi^2/3)\,k_B/|e|$. The quantitative agreement found below 40 K between this calculation and the measured Nernst coefficient (black symbols) lends strong evidence that the enhanced Nernst signal originates from the significant mobility mismatch across the sample. The red dashed line displays the Hall mobility $\mu_H(T)$. $\nu(T)$ is expected to approximately scale with $\mu_H(T)$ assuming a weak temperature dependence of the latter, which indeed holds above 40 K (see text for details).

## Acknowledgements

We thank J.L. Snyman for valuable assistance with materials synthesis, and J.L. Luo, N.L. Wang, Y.F. Yang, H. Z. Zhao, M. Baenitz, and G. Kotliar for discussions. P.S., B.W. and J.Z. are thankful to the financial support from the MOST of China (Grant Nos: 2012CB921701 and 2015CB921303), the National Science Foundation of China (Grant No: 11474332), and the Chinese Academy of Sciences through the strategic priority research program (Grant No: XDB07020200). A.M.S. thanks the SA-NRF (93549) and the FRC/URC of UJ for financial assistance, and the IOP-CAS for a visiting professor fellowship. J.M.T. thanks COST Action MP1306 EUSpec for support. M. S. and B. B. I. thank the Danish National Research Foundation (DNRF93) for support.

## Author contributions

P.S. and F.S. conceived the idea. P.S., B.W., and J.Z. performed all the electrical and thermal transport experiments and analyzed the data, after discussion with other authors. J.M.T performed the electronic structure and thermoelectric calculations. A.M.S., M.S. and B.B.I. prepared the sample and carried out sample characterization. P.S. and F.S. wrote the manuscript and all authors commented on and edited the manuscript.